\documentclass[12pt,twoside]{article}

\usepackage{amsmath, amsthm, amssymb}

\date{}

\begin{document}

\centerline{}

\centerline{}

\centerline {\Large{\bf Proving the Relativistic Rotation Paradox}}

\centerline{}

\newcommand{\mvec}[1]{\mbox{\bfseries\itshape #1}}

\centerline{\bf {Christian P. H. Salas}}

\centerline{}

\centerline{Faculty of Mathematics \& Statistics, The Open University, Milton Keynes} 
\centerline{and Department of Mathematics, Waltham Forest College, London}
\centerline{e-mail: c.p.h.salas@open.ac.uk}

\centerline{}

\newtheorem{Theorem}{\quad Theorem}[section]

\newtheorem{Definition}[Theorem]{\quad Definition}

\newtheorem{Corollary}[Theorem]{\quad Corollary}

\newtheorem{Lemma}[Theorem]{\quad Lemma}

\newtheorem{Example}[Theorem]{\quad Example}

\centerline{}\bigskip

\centerline{\bf {Abstract}}\bigskip

\textit{An apparent paradox in Einstein's Special Theory of Relativity, known as a Thomas precession rotation in atomic physics, has been verified experimentally in a number of ways. However, somewhat surprisingly, it has not yet been demonstrated algebraically in a straightforward manner using Lorentz-matrix-algebra. Authors in the past have resorted instead to computer verifications, or to overly-complicated derivations, leaving undergraduate students in particular with the impression that this is a mysterious and mathematically inaccessible phenomenon. This is surprising because, as shown in the present note, it is possible to use a basic property of orthogonal Lorentz matrices and a judicious choice for the configuration of the relevant inertial frames to give a very transparent algebraic proof. It is pedagogically useful for physics students particularly at undergraduate level to explore this. It not only clarifies the nature of the paradox at an accessible mathematical level and sheds additional light on some mathematical properties of Lorentz matrices and relatively-moving frames. It also illustrates the satisfaction that a clear mathematical understanding of a physics problem can bring, compared to uninspired computations or tortured derivations.}\bigskip


\centerline{}

\section{Introduction}
Multiplying two Lorentz boosts whose velocity vectors are collinear gives a third Lorentz boost whose velocity can be calculated from the first two using the Einstein velocity addition law. If the two original velocities are not collinear, however, we do not get a pure Lorentz boost as the product, but rather a Lorentz boost multiplied by a certain $4 \times 4$ matrix whose columns and rows are orthonormal. This orthogonal matrix has the effect of rotating the spatial components of vectors in spacetime, while leaving their temporal component unaffected. 

An interesting review paper written by the eminent British mathematician I. J. Good \cite{Good} discusses this relativistic rotation paradox, but Good clearly struggles to provide a straightforward algebraic proof in the 3+1 case. Mathematically, it is necessary to show in a four-dimensional Minkowski spacetime that a certain matrix product involving two Lorentz boosts with linearly independent velocity vectors is generally equivalent to an orthogonal Lorentz matrix. In section 8 of \cite{Good}, which seeks to prove the paradox `beyond any doubt', Good admits to being unable to do this algebraically and instead resorts to providing numerical confirmations. Such computational confirmations are easy to carry out, so he suggests that `only a short and elegant algebraic proof would be worthwhile'. One of the motivations for the present note is that a proof like this still seems to be lacking in the literature. 

What we have at present in the way of mathematical demonstrations are either elaborate approximations involving power series and extraneous assumptions such as infinitesimally small relative velocities (see, e.g., section 7.3 in \cite{Goldstein}), or otherwise lengthy and overly-sophisticated expositions usually not easily accessible to, say, undergraduate physics students. Some expositions are intended to be more accessible but still seem rather involved and/or do not make use of Lorentz-matrix-algebra, e.g., \cite{Ferraro} and \cite{ODonnell}. 

Due to the lack of a short and transparent algebraic treatment, this interesting relativistic phenomenon is simply left unmentioned and unexplored in almost all undergraduate texts, which seems a pity. The following argument could be used shortly after introducing Lorentz transformation matrices and their properties to budding relativists.  
\section{Proof of the rotation paradox}
Let $G = \text{diag}(1, -1, -1, -1)$ be the metric tensor in a four-dimensional Minkowski manifold with events specified by a time coordinate $x^0 = ct$ and rectangular spatial coordinates $x^1, x^2, x^3$. A $4 \times 4$ Lorentz matrix $\Lambda$ preserves the quadratic form $x^T G x$ in the sense that if $y = \Lambda x$ then $y^T G y = x^T G x$, so
\begin{equation}
\Lambda^T G \Lambda = G
\end{equation}
The set of all Lorentz matrices thus defined constitutes a group under matrix multiplication, so inverses and products of Lorentz matrices are also Lorentz.  

Let $O$, $\overline{O}$ and $\overline{\overline{O}}$ be three inertial frames with collinear axes and with their origins initially coinciding. Let $\beta = (\beta_i) = \big(\frac{v_i}{c}\big)$, $i = 1, 2, 3$, be the $3 \times 1$ velocity vector of $\overline{O}$ relative to $O$ with corresponding Lorentz factor $\gamma = \frac{1}{\sqrt{1 - \beta^2}}$, where $\beta^2 \equiv \beta^T \beta$. Similarly, let a vector $\overline{\beta} = (\overline{\beta}_i)$, which is not collinear with $\beta$, be the velocity vector of $\overline{\overline{O}}$ relative to $\overline{O}$ with corresponding Lorentz factor $\overline{\gamma}$. Using a standard formula, e.g., formula (24) in \cite{Good} or formula (2.59) in \cite{Moller}, the velocity vector of $\overline{\overline{O}}$ relative to $O$ is given by 
\begin{equation}
\overline{\overline{\beta}} = \frac{\overline{\beta} + \beta[\gamma + (\gamma - 1)(\overline{\beta}^T\beta)/\beta^2]}{(1 + \overline{\beta}^T\beta)\gamma}
\end{equation} 
with corresponding Lorentz factor $\overline{\overline{\gamma}}$. Using a simplification similar to one described in section 7.3 of \cite{Goldstein}, we can let the plane defined by the vectors $\beta$ and $\overline{\beta}$ be the $\overline{x}^1\overline{x}^2$-plane of $\overline{O}$ so that $\overline{\beta}_3 =0$, and we can arrange the frames $O$ and $\overline{O}$ so that the vector $\beta$ is along the $x^1$ axis of $O$, implying $\beta_2 = \beta_3 = 0$. We can do this for any given pair of velocity vectors which are not collinear, so there is no loss of generality here. Then (2) gives
\begin{equation}
\overline{\overline{\beta}}_1 = \frac{\overline{\beta}_1 + \beta_1}{1 + \overline{\beta}_1 \beta_1}
\end{equation}
\begin{equation}
\overline{\overline{\beta}}_2 = \frac{\overline{\beta}_2}{(1 + \overline{\beta}_1 \beta_1) \gamma}
\end{equation}
\begin{equation}
\overline{\overline{\beta}}_3 = 0
\end{equation}
and a standard formula, e.g., formula (7.11) in \cite{Goldstein}, allows us to write Lorentz transformations $L$ and $\overline{L}$ from $O$ to $\overline{O}$ and from $\overline{O}$ to $\overline{\overline{O}}$ respectively as    
\begin{equation}
L = 
\begin{pmatrix}
\gamma & -\gamma \beta_1 & 0 \ & 0 \ \\
-\gamma \beta_1 & \gamma & 0 \ & 0 \ \\
0 & 0 & 1 \ & 0 \\
0 & 0 & 0 \ & 1 
\end{pmatrix}
\end{equation}
and
\begin{equation}
\overline{L} = 
\begin{pmatrix}
\overline{\gamma} \ \ & -\overline{\gamma} \overline{\beta}_1 \ \ & -\overline{\gamma} \overline{\beta}_2 \ \ & 0 \ \\
-\overline{\gamma} \overline{\beta}_1 \ \ & 1+(\overline{\gamma}-1) \frac{\overline{\beta}_1^2}{\overline{\beta}^2} \ \ & (\overline{\gamma}-1) \frac{\overline{\beta}_1\overline{\beta}_2}{\overline{\beta}^2} \ \ & 0 \ \\
-\overline{\gamma} \overline{\beta}_2 \ \ & (\overline{\gamma}-1) \frac{\overline{\beta}_1\overline{\beta}_2}{\overline{\beta}^2} \ \ & 1+(\overline{\gamma}-1) \frac{\overline{\beta}_2^2}{\overline{\beta}^2} \ \ & 0 \\
0 \ \ & 0 \ \ & 0 \ \ & 1 
\end{pmatrix}
\end{equation}
A Lorentz boost $\overline{\overline{L}}$ from $O$ to $\overline{\overline{O}}$ with velocity vector $\overline{\overline{\beta}}$ would be a matrix like (7), but with $\overline{\overline{\gamma}}$, $\overline{\overline{\beta}}_1$, $\overline{\overline{\beta}}_2$ and $\overline{\overline{\beta}}$ replacing $\overline{\gamma}$, $\overline{\beta}_1$, $\overline{\beta}_2$ and $\overline{\beta}$ respectively. 

The relativistic rotation paradox is that, in general, $\overline{\overline{L}} \neq \overline{L} \times L$, but rather
\begin{equation}
\overline{\overline{L}} = R \times \overline{L} \times L
\end{equation}  
or equivalently
\begin{equation}
R = \overline{\overline{L}} \times L^{-1} \times \overline{L}^{-1}
\end{equation}  
where (9) is equation (28) in \cite{Good}. Numerical evidence in \cite{Good} suggests that
\begin{equation}
R = 
\begin{pmatrix}
1 & 0 & 0  & 0  \\
0 & r_1 & s_1  & t_1  \\
0 & r_2 & s_2 & t_2 \\
0 & r_3 & s_3 & t_3 
\end{pmatrix}
\end{equation}
where the $3 \times 3$ submatrix in (10) is orthogonal. An approximation to (10) is also provided in equation (7.21) of \cite{Goldstein} under the assumptions that the components of $\overline{\beta}$ are small and only need to be retained to first order, that $\overline{\gamma} \approx 1$, and that the distinction among $\gamma$, $\overline{\gamma}$ and $\overline{\overline{\gamma}}$ can be ignored to first order. 

However, it is straightforward to obtain an exact algebraic proof that $R$ in (9) is indeed an orthogonal matrix of the type given in (10) by observing that $R$ must be Lorentz, since it is a product of Lorentz matrices. Therefore all that is required to prove the rotation paradox is to show that the 00-element of $\overline{\overline{L}} \times L^{-1} \times \overline{L}^{-1}$ is equal to $1$, and that all the remaining elements in the first row are equal to zero, since any Lorentz matrix with a first row of this form must necessarily be an orthogonal matrix of the type given in (10). This assertion can easily be verified by substituting a generic $4 \times 4$ matrix with first row of the form $(1 \ 0 \ 0 \ 0)$ into the left-hand side of (1), setting the result equal to $G$ on the right-hand side, and then comparing corresponding elements.  

Note that $L^{-1}$ and $\overline{L}^{-1}$ are immediately obtained from (6) and (7) simply by removing the negative signs in the first row and first column. To prove that the 00-element of $\overline{\overline{L}} \times L^{-1} \times \overline{L}^{-1}$ equals 1, multiply the first row of $\overline{\overline{L}}$ by each of the columns of $L^{-1}$ to get the $1 \times 4$ row vector
 \begin{equation}
\begin{pmatrix}
\overline{\overline{\gamma}} \gamma (1 - \overline{\overline{\beta}}_1 \beta_1) \ \ & \overline{\overline{\gamma}} \gamma (\beta_1 - \overline{\overline{\beta}}_1) \ \ & -\overline{\overline{\gamma}} \ \overline{\overline{\beta}}_2 \ \ & 0   
\end{pmatrix}
\end{equation}
and then multiply this row vector by the first column of the matrix $\overline{L}^{-1}$ to get
\begin{equation*}
\gamma \ \overline{\gamma} \ \overline{\overline{\gamma}}(1 - \overline{\overline{\beta}}_1 \beta_1 + \overline{\beta}_1 \beta_1 - \overline{\beta}_1 \overline{\overline{\beta}}_1) - \overline{\gamma} \ \overline{\overline{\gamma}} \ \overline{\beta}_2 \overline{\overline{\beta}}_2
\end{equation*}
\begin{equation*}
= \frac{(1 + \beta_1 \overline{\beta}_1)^2 - (\beta_1 +  \overline{\beta}_1)^2 -  \overline{\beta}_2^2(1 - \beta_1^2)}{\sqrt{1 - \beta_1^2}\sqrt{1 - \overline{\beta}_1^2 - \overline{\beta}_2^2}\sqrt{(1 + \beta_1 \overline{\beta}_1)^2 - (\beta_1 +  \overline{\beta}_1)^2 -  \overline{\beta}_2^2(1 - \beta_1^2)}} = 1
\end{equation*}
as required. To prove that the 01-element of $\overline{\overline{L}} \times L^{-1} \times \overline{L}^{-1}$ equals 0, multiply the row vector in (11) by the second column of $\overline{L}^{-1}$ to get
\begin{equation*}
\gamma \ \overline{\gamma} \ \overline{\overline{\gamma}}(\overline{\beta}_1 - \beta_1 \overline{\beta}_1 \overline{\overline{\beta}}_1) + 
\bigg[1+(\overline{\gamma}-1) \frac{\overline{\beta}_1^2}{\overline{\beta}^2}\bigg] 
\gamma \ \overline{\overline{\gamma}} \ (\beta_1 - \overline{\overline{\beta}}_1) 
- \overline{\overline{\gamma}} \ \overline{\overline{\beta}}_2 \bigg[(\overline{\gamma}-1) \frac{\overline{\beta}_1\overline{\beta}_2}{\overline{\beta}^2}\bigg]
\end{equation*}
\begin{equation*}
= \frac{\overline{\beta}_1(1 - \beta_1^2)}{(1 - \beta_1^2)(1 - \overline{\beta}_1^2 - \overline{\beta}_2^2)} - 
\frac{\overline{\beta}_1(\overline{\beta}_1^2 + \overline{\beta}_2^2)}{(\overline{\beta}_1^2 + \overline{\beta}_2^2)(1 - \overline{\beta}_1^2 - \overline{\beta}_2^2)} = 0
\end{equation*}
as required. To prove that the 02-element of $\overline{\overline{L}} \times L^{-1} \times \overline{L}^{-1}$ equals 0, multiply the row vector in (11) by the third column of $\overline{L}^{-1}$ to get
\begin{equation*}
\gamma \ \overline{\gamma} \ \overline{\overline{\gamma}}(\overline{\beta}_2 - \beta_1 \overline{\beta}_2 \overline{\overline{\beta}}_1) +   \bigg[(\overline{\gamma}-1) \frac{\overline{\beta}_1\overline{\beta}_2}{\overline{\beta}^2}\bigg]\gamma \ \overline{\overline{\gamma}} \ (\beta_1 - \overline{\overline{\beta}}_1) -
\overline{\overline{\gamma}} \ \overline{\overline{\beta}}_2\bigg[1+(\overline{\gamma}-1) \frac{\overline{\beta}_2^2}{\overline{\beta}^2}\bigg] 
\end{equation*}
\begin{equation*}
= \frac{\overline{\beta}_2(1 - \beta_1^2)}{(1 - \beta_1^2)(1 - \overline{\beta}_1^2 - \overline{\beta}_2^2)} - 
\frac{\overline{\beta}_2(\overline{\beta}_1^2 + \overline{\beta}_2^2)}{(\overline{\beta}_1^2 + \overline{\beta}_2^2)(1 - \overline{\beta}_1^2 - \overline{\beta}_2^2)} = 0
\end{equation*}
as required. Finally, to prove that the 03-element of $\overline{\overline{L}} \times L^{-1} \times \overline{L}^{-1}$ equals 0, multiply the row vector in (11) by the fourth column of $\overline{L}^{-1}$. This equals 0 by inspection, so the relativistic rotation paradox is proved. 

\centerline{}

\end{document}